\documentclass[12pt, a4paper]{article}
\usepackage{pdfpages}
\usepackage{amsmath,amssymb,amsthm}
\pdfoutput=1

\newcommand{\beq}{\begin{equation}}
\newcommand{\eeq}{\end{equation}}
\newcommand{\beqa}{\begin{eqnarray}}
\newcommand{\eeqa}{\end{eqnarray}}

\title{On the asymptotics of AKT statistics}
\author
{L\'{\i}dia Rejt\H{o}$^{1}$
\and
G\'abor Tusn\'ady$^{2}$}

\date{2013. June 7:14:45 -- September 15, 2013}

\begin{document}

\maketitle
\begin{abstract}
In our days there is a widespread analysis of Wasserstein distances between theoretical and empirical measures.
One of the first investigation
of the topic is given in the paper written by Ajtai, Koml\'os and Tusn\'ady in $1984.$ Interestingly
all the neighboring questions posed by that paper were settled already without the original one. 
In this paper we are going to delineate the limit behavior of the original
statistics with the help of computer simulations.
At the same time we 
kept an eye on theoretical grasping of the problem.
Based on our computer simulations our opinion is that  the limit distribution is Gaussian.
\end{abstract}

\footnotetext[1]{Department of Applied Economics and Statistics, University of Delaware, Newark, Delaware, USA \\rejto@udel.edu}
\footnotetext[2]{Alfr\'ed  R\'enyi
Mathematical Institute of the Hungarian Academy of Sciences,
Budapest, Hungary\\ tusnady.gabor@renyi.mta.hu}

\bigskip

\section{Introduction}
Let $((X_i,Y_i),\; i=1,\ldots,n)$ be independent points uniformly distributed on the $d$ dimensional unit cube.
The AKT statistic $W_n$ is the minimum of
\beq {\label {eq:match}}
\sum^n_{i=1} \mid X_i - Y_{\pi(i)}\mid^2
\eeq
taken on all permutations $\pi$ of $(1,\ldots,n).$ We call the statistics Wasserstein distance.
Using computer experiments 
we are going to investigate the asymptotic behavior of $W_n$  as $d=2$ and $n$ goes to infinity. 
Based on our computer simulations we think that if $n\leq 2048,$ then
the hazard rate function of the distribution of $W_n - \beta \log(n)$ is proportional to
an appropriately scaled Gaussian distribution function with $\beta = 0.11.$

\bigskip

\section{History}

The AKT statistics were introduced in  \cite{AKT}.
In that paper the authors proved that there exist positive constants $c_1,c_2$ such that with probability 
going to 1 
\beq{\label{eq:orig}}
c_1\log(n)<W_n<c_2\log(n)
\eeq 
holds true. The origin of this statistic goes back to the one dimensional
Shapiro--Wilks statistic for testing normality.
In \cite{BGM} the case $d=1$ was investigated.
The AKT statistics were generalized in  \cite{Talp} while
a new proof was given for (\ref{eq:orig}) in  \cite{Talb}. 
Talagrand next investigated the case $d\geq 3$ in \cite{Talo}
and settled all the observable questions.
Talagrand's estimates for $d=2$ were
sharpened in \cite{PM}. A broad introduction to  transportation theory is presented in \cite{Vil}.
The first application of the AKT theorem was in bin packing \cite{CGJ}. A numerical investigation
of the distribution of the statistic was given in \cite{Nos}. 
Matchings on the Euclidean ball were investigated in \cite{BO}. The paper
\cite{Torri} investigates the rate of convergence in abstract settings.
An investigation of different types of matchings  was given  in \cite{HPPS}.
In \cite{CR} distributions supported on a manifold embedded in a Hilbert space were investigated.

\bigskip

\section{Dyadic dynamics}

Starting with $n=2^0$ points we evolve $n=2^k, k=1,2..$ point pairs in such a way
that in each step the original points  remain 
 and simultaneously a new set of point pairs are supplanted.
We call the original kids "old" and the new ones "young". Having 
two types of kids, boys and girls, there are four 
possibilities for matching:

-- old girls with old boys ($OO_k$),

-- old girls with young boys ($OY_k$),

-- young girls with old boys ($YO_k$),

-- young girls with young boys ($YY_k$).

The matchings $OO_k$ and $YY_k$ are independent, and similarly $OY_k$ and $YO_k$ are
independent while the costs $OO_k$ and $OY_k$ are correlated.
The cost is the Wasserstein distance $W_n$ between the corresponding set of points.
For small values of $n$ the approximation $$EW_n = \beta \log(n+\alpha) + \gamma$$
is more accurate. Figure 1 shows the
averages of costs of  $296$ repetitions up to $k=11$ with the above
theoretical function. Here $\alpha=0.379, \beta=0.160, \gamma=0.117.$
The densities of the
normalized differences of $W_n - (\beta \log(n+\alpha) + \gamma)$ are
given in Figure 2. Next generation costs $OO_{k+1}$ are close to the
averages $$(OO_k+OY_k+YO_k+YY_k)/4,$$ and the variance of the error is $0.101.$
The quadratic recursion
\beq{\label{eq:dynam}}
OO_{k+1} = (OO_k+OY_k+YO_k+YY_k)/4 + V_{k+1}
\eeq
with an iid noise $V_{k+1}$ agrees with our conjecture indicating
that the quadratic recursion might be valid  for two dimensions only.

In Euclidean space for arbitrary points $a,b,c,d$
\beq {\label{eq:Euc}}
\mid (a+b) - (c+d) \mid^2 = \mid a-c\mid^2 + \mid a-d\mid^2 + \mid b-c\mid^2 + \mid b-d\mid^2 - \mid a-b\mid^2 - \mid c-d\mid^2
\eeq
holds true. Similarly in our case
\beq {\label{eq:six}}
OO_{k+1} = a*(OO_k+OY_k+YO_k+YY_k) - b*(GG_k+BB_k) + V_{k+1}
\eeq
holds true -where $GG$ and $BB$ are the girls-girls, boys-boys distances. 
For stacionarity 
\beq{\label{eq:stac}}
4*a-2*b=1
\eeq 
must be hold.
According to our computer experiences
$a=0.45,\; b=0.41$ and the error term is $0.06.$ The distribution of the $V$ error term seems to be double exponential.

\bigskip

\section{Pictorial presentation of the Hungarian Algorithm}

The engine of the Hungarian Algorithm is the system of shadow
prices $a(i), b(j)$ such that
$$
b(j) - a(i) \leq c(i,j),
$$
where $c(i,j)$ is the cost of the marriage of a pair $(i,j).$
In Figures 3 and 4 we give a
pictorial presentation of the prices
for $n=1024$ and $n=2048.$ Here we 
used toroid distances 
and the colors
of points of the unit square shows the price of the closest
kid to the point, i.e. the $a(i)$ in case the closest kid is a girl
and the $b(j)$ otherwise.
The keys of colours are given on the margin.
Pixel statistics are given for the colours, first number
stands for wives, second for husbands. E.g.
in Figure 3 there are  $32$  yellow wives and $69$ husbands.
Dark blue represents the minimal value, which is zero for
the value of the last wife is always zero.
The maximal value for $n=1024$ is 0.02499 and it is 0.01654
for $n=2024$, they are represented by yellow colour.
The kids themselves are complementarily coloured for seeing them.
Circles represent
girls, squares represent boys. Marriages are shown by gray lines. For
neglecting complications of toroid topology girls having a husband
on the other side of the unit square are marked with an Andrew cross
while their husband is not assigned. Interestingly usually there is one
lake and one hill.

The actual total costs of the marriages are
$W_{1024}=0.996, W_{2048}=1.654,$ thus the average distance in a
marriage is around $0.02$ causing the same or neighboring
colours for wife and husband. We know that for a married couple $(i,j)$
we have
$$
b(j) - a(i) = c(i,j),
$$
hence
all marriages are upwards directed according their colour.
 $1024$ point-pair
from the $2048$ point-pairs in Figure 4 are identical with point-pairs
given in Figure 3, the others are generated independently.
The situation not much changed.
The bordering lines of colours for $n=1024$ are simpler than for $n=2024.$

\bigskip

\section{Asymptotics}

For the distribution of $W_n$  the best approximation achieved  by us  is a distribution determined by its
hazard rate function. 
Let us denote the tail distribution of a one--dimensional random variable by $Q(t).$
The hazard rate $r(t)$ is the derivative of $-\log Q(t).$ Customarily hazard rates are defined for survival functions,
belonging to positive random variables. 
The concept is extendable naturally for random variables taking values between
$-\infty$ and $\infty$ too. Although $W_n$ is positive, the best approximation comes from this broader class
of distributions. The hazard rate is an arbitrary non-negative function having infinite integral on the
real line. In our case it is proportional to  $\Phi(t)$ the distribution function of a standard normal random variable. 
Its integral up to an $x$ is  equal to $x \Phi(x)+\varphi(x),$ where $\varphi(x)$ is the standard normal density function. 
By definition it equals to $-\log Q(x)$. Applying the relation
$r(x)=f(x)/Q(x),$ where $f(x)$ stands for the density function we get
\beq {\label{eq:basic}} 
f(x) = \Phi(x)\exp(-x\Phi(x)-\varphi(x)).
\eeq
The derivative of $\log f(x)$ is $\varphi(x)/\Phi(x) - \Phi(x),$ which is for $x$ going to $-\infty$ close to $-x,$ 
for $x$ going to $\infty$ it is going to $(-1)$ and it is monotone decreasing. Thus for negative $x$-s the distribution
resembles a double exponential distribution but for positive $x$-s it turns to be a single exponential.
That is why $f$ is  increasing first sharply to its modus around $0.3$ and it is slowly decreasing after that value.
  
We are going to build up a three parameter family around the above distribution. The first step is Cox regression.
Let $\lambda$ be arbitrary positive
number,
 then $\rho_\lambda (t) = \lambda \Phi (t)$ is the hazard rate of the
distribution with tail $Q(t)^\lambda$ and for any real $\mu$ and positive $\sigma$ the linear transformation
presents the density
\beq {\label{eq:three}}
f(x\mid \mu,\sigma,\lambda) = {\frac{\lambda}{\sigma}} \Phi (y) \exp(-\lambda (y\Phi(y)-\varphi(y)), \quad {\text {where}}\quad y=(x-\mu)/\sigma.
\eeq
Let us denote by $X_\lambda$ a random variable corresponding to the hazard rate $\varrho_{\lambda}.$
Let $\lambda>0, \sigma>0, \mu$ be arbitrary real numbers, then
$f(x\mid \mu,\sigma,\lambda)$ is the density of $\sigma X_\lambda + \mu.$ 
Thus $\mu$ is the location parameter,
$\sigma$ is the scale parameter and $\lambda$ is the shape parameter of the distribution. 
Figure 5. shows the density of the distribution of $X_\lambda/(\lambda)^\alpha$ with  $\alpha=0.4$, for eleven different
$\lambda$  from $0.1$ to $10.$
For small $\lambda$,  one can see that the distribution of $X_\lambda/(\lambda)^\alpha$
has an intensively increasing
 first phase and after  it becomes similar to
the exponential distribution. On the other hand for large $\lambda$ it turns to be just the opposite. 

The estimates of the parameters $\sigma$ and $ \lambda$ are highly correlated, 
unfortunately our present sample size is not large
enough to decide whether the value of the parameter $\lambda$ differs significantly from $1$
or not. If it does so then the question is the tendency of $\lambda$ as $n$ goes to infinity. 
 With $\lambda=1$  the tendency of $\sigma$ is not clear: it is
increasing from $0.08$ up to $0.14$ and perhaps remains bounded.

The dynamic (\ref{eq:six}) has a second condition (not presented here)
in addition to (\ref{eq:stac}).  
It ensures the boundedness of variance.
For that condition we have to know the covariances of the costs of different matchings. First it was an unsettled riddle
for us what is the joint distribution for the six transportation 
 costs plus error term resulting in a Gaussian hazard rate with
stationary distribution. 
 In order to settle this question
the ideas of the paper \cite{RS} turned out to be useful: 
even the reference to the Euclidean relation (\ref{eq:Euc}) comes from the possibility that Euclidean relations 
might be generalized into transportation  equations. In the next section we are going to discuss a partial solution of
the riddle.

\bigskip

\section{A seven dimensional model}

For fixed $k$ we generate four times
 $2^k$ of points,
 two sets of girl and two sets of boys. Let us label then $A,B,C,D$. 
($A$ is the set of old girls, $B$ is the set of young girls, $C$ is the set of old boys, and $D$ is the set of young boys.)
There are six Wasserstein distances among them:
$$
W_1=(A,C), \; W_2=(A,D), \; W_3=(B,C), \; W_4=(B,D), \; W_5=(A,B), \; W_6=(C,D).
$$ 
$W_1$ is independent of $W_4$ and its relation with the others is symmetric. In building a joint distribution for the six
variables we use an autoregressive model: the conditional distribution of $W_{i+1}$  given  $(W_1,\ldots,W_i),$ is a distribution
with Gaussian hazard rate with arbitrary $\sigma_i,$ $\lambda=1$ and
$$
\mu_i = \gamma_{i0} + \sum^i_{s=1} \gamma_{is} W_s.
$$
The pairs $(\gamma_{i0},\mu_i)$ for $k=10$ are the followings:
$$
 (1.142,  0.133) \;\;
 (0.907,  0.129) \;\;
 (0.923,  0.127) \;\;
 (0.706,  0.125) \;\;
 (0.553,  0.117) \;\;
 (0.591,  0.117).
$$
For smaller $k$ the tendency is similar, the leading terms $\gamma_{i0}$ follow the general logarithmic trend and the $\sigma_i$-s are
practically the same. It goes without saying that they diverse in $i$. The autoregressive coefficients are the following:

\newpage

\bigskip
\begin{center}
{\bf {Table 1.}}
\end{center}

\begin{tabular}{lrrrr}
$\gamma_{21} = 0.194$&&&&\\
$\gamma_{31} = 0.176 $  &  $ \gamma_{32} = 0.006$&&&\\
$\gamma_{41} = -0.040$  & $\gamma_{42} = 0.188$   &    $\gamma_{43} = 0.204$&&\\
$\gamma_{51} = 0.101$  &  $\gamma_{52} = 0.140 $ & $    \gamma_{53} = 0.126$&    $ \gamma_{54} = 0.123$&\\
$\gamma_{61} = 0.126 $ & $ \gamma_{62} = 0.139$ & $     \gamma_{63} = 0.138 $ & $\gamma_{64} = 0.108$ & $   \gamma_{65} = -0.054$
\end{tabular}

\bigskip

These coefficients are mostly small and they are negative for independent pairs. 
The conditional distribution of $\tilde W_1$ for $k+1$ is similar: 
the gammas follow the original pattern found with linear regression and the standard deviations are around $0.055.$ 

We can test the model in the following way. Having a well parametrized $67$ dimensional distribution we can generate independent
$67$ dimensional random vectors as many times
as many samples we have in the unit square. Presently it is $817$. Next
we use the Hungarian algorithm to determine the Wasserstein distance of the two point systems: it is around $1440$.
 Finally we generate new random samples and determine their distances. Interestingly it is {\it {larger}}  then $1440$, its average
is $1470$ with deviance $10$. It means that the structure of the unit-square sample is a bit tighter than our autoregressive
scheme. But a simple trick settles the trouble: if we multiply all standard deviation parameter $\sigma$ with $0.975$ then
the unit-square Wasserstein get in the middle of model Wassersteins. 

If $X$ is an arbitrary real random variable and $Q(x)$ is its tail probability $P(X>x),$ then the distribution of $Q(X)$ is uniform
in the interval $(0,1).$ The integrated hazard rate $$R(x)=\int^x_{-\infty}r(t)dt$$ 
equals to $-\log(Q(x)),$ thus the distribution of $R(X)$ is
standard exponential and the distribution of $\exp(-R(X))$ is again uniform in $(0,1)$. So we can test the hypothesis that
the distribution of the $67$ dimensional Wasserstein statistics belongs to the three parameter family (\ref{eq:three})
testing the uniformity of this statistics. 
Of course we can not use the original Kolmogorov cut-points because we use estimated
parameters, we have to calibrate the cut-points by random sample. In our case 
for levels $0.05, 0.01$ they are $0.84, 1.37.$
The value of the actual statistic is $1.00$ 
since this new Wassertstein statistics might have Gaussian hazard rate.
The corresponding statistics for points in the unit square are the followings.

\newpage

\bigskip
\begin{center}
{\bf{Table 2.}}
\end{center}
{\hspace{2cm}{
\begin{tabular}{lrcccccc}
 $k$   &  No   &    Kolmogorov     &    R1        &     R2       &   R3     &   R4     &    R5      \\
 0   & 4902      &   3.0578          &   0.6148     &    0.6036    &  0.5354  &  0.6367  &   0.3776   \\  
 1   & 4902      &   1.8130          &   0.7537     &    0.8154    &  0.7425  &  0.6165  &   0.5595   \\
 2   & 4902      &   1.2752          &   0.6765     &    3.6339    &  0.6176  &  0.7559  &   0.3969   \\
 3   & 4902      &   1.1030          &   0.4839     &    0.9168    &  0.5846  &  0.6992  &   5.7871   \\
 4   & 4902      &   0.9454          &   0.9077     &    0.8388    &  0.7162  &  0.5211  &   0.7878   \\
 5   & 4902      &   1.1828          &   0.5653     &    0.6978    &  0.7649  &  0.5019  &   0.5467   \\
 6   & 4902      &   1.2998          &   0.5306     &    0.5985    &  0.7461  &  0.5844  &   0.9778   \\
 7   & 4902      &   0.9876          &   0.8416     &    0.7448    &  0.5140  &  0.6265  &   0.7459   \\
 8   & 4902      &   0.9327          &   0.7223     &    0.7323    &  0.6713  &  0.9064  &   0.5782   \\
 9   & 4902      &   0.9697          &   0.8291     &    0.6565    &  0.9028  &  0.5890  &   0.7438   \\
10   & 4902      &   1.0793          &   0.6249     &    0.6609    &  0.7140  &  0.5499  &   0.6290   \\
11   &  817      &   0.6508          &   0.8411     &    0.7843    &  1.0557  &  1.1890  &   0.9961   \\
\end{tabular}
}}
\bigskip

Here $k$ is number of number of doublings in the dynamics, 
No is the sample size. We have $817$ different runs and for $k=11$ 
and the sample size  for $k<11$ it is $6*817$. 
We multiply the maximal difference between empirical and theoretical distributions
with $\sqrt {{\mbox{No}}}$ but for $k<11$ the Wasserstein distances are not independent. It might be the
reason that even for $k=10$ we have a borderline result. Of course we are aware the fact that for $k=0$ 
the Wasserstein statistics has a different
distribution and perhaps the situation is similar for small $k$-s. This tendency is clearly shown by the Kolmogorov statistics.

We generated five times data matrix by the theoretical model.
 Columns headed by R1,...,R5 gives the corresponding Kolmogorov
statistics. As one can observe the statistics have rather long tail arising perhaps from the interwoven dependence structure.

\bigskip

\section{The Ajtai statistics}

In paper \cite{AKT} the saddle point method is used to prove inequality (\ref{eq:orig}). Appropriate
Lipschitzean functions are developed to prove the left hand side 
while for the right hand side an ad-hoc matching algorithm due to Mikl\'os Ajtai is used.
The algorithm is based on the following elementary concept.

{\bf{Definition}}\quad Let $s$ be a positive integer, $t=2s$ and $A=(a_1,\ldots,a_t)$ arbitrary reals. The median bits 
$B=(b_1,\ldots,b_t)$ of $A$ are $(0,1)$-s defined by the properties
$$
\sum^t_{k=1}b_k = s; \quad {\text{if}} \;((b_i=0) \;{\text{and}}\; (b_j=1)), \; {\text{then}} \; a_i \leq a_j.
$$
In case there are no ties in $A,\; B$ is uniquely defined. 

Let $k$ be an arbitrary natural number, $n=4^k,$ and $Z=(Z_i,\; i=1,\ldots,n)$ be a system of arbitrary two-dimensional points.
In a step-by-step procedure we order a $2k$ long bit sequence to each of  the points in $Z.$ As an initial step we  construct a set $A$ from
the first coordinates of $Z$. Having the corresponding $B$ we divide $Z$ into two subsets according to the bits in $B$
and for each of that subsets we form the median bits from the second coordinates.

From these initializations we proceed in the same manner. Using the bit sequences ordered to the points so far first
we develop the next bits from the subsets formed of the first coordinates having the same bit sequences generated so far and
next we turn to the second coordinates. Applying the procedure independently for two iid two dimensional 
sets $X$ and $Y$ in the role 
 of  $Z$ the matching is supplied by the identical bit sequences. In course of the algorithm step by step
the size of subsets is halved and finally it is reduced to one, thus for all possible $2k$ long bit sequence
we have exactly one point in $X$ and one in $Y$, and it makes the matching. In a certain sense we construct a two-dimensional
ordering merging the orders according the two coordinates.
 Let $b$ be an arbitrary $2k$ long bit sequence and
 $$
c = \sum^k_{i=1} b_{2i-1}, \quad d = \sum^k_{i=1} b_{2i}.
$$ 
The expected value of the first coordinate of the corresponding points
in $X$ or $Y$ are labelled by $b$ 
is $c/(2^k+1)$
and for the second coordinate it is $d/(2^k+1).$ (Let us note here that in our notation $X$ and $Y$ are two sets of {\it two dimensional points}
and the coordinates are $(x_{i1},x_{i2}), \; (y_{i1},y_{i2})).$

Let us observe that the marginal quantile transform applies for the algorithm: the matching itself does not depend on any monotone transform.
(The marginal quantile transform is $F(x_{i1}),\;i=1,\ldots,n$ for the first coordinates and $G(x_{i2}),\;i=1,\ldots,n$ for
the second coordinates if the coordinates are independent and the distribution of the first coordinates  if $F$ and that
is $G$ for the second coordinates.)
Utilizing the mean limits offered by the algorithm we are deeply concerned that the limit distribution for any distributions
has the form
$$
\sum^n_{i=1} c_i x_i^2
$$
where the $x_i$-s are independent standard normals and the $c_i$-s are appropriate coefficients. We guess that such random
variables have monotone increasing hazard rates, and for certain coefficients the hazard rate function is bounded while for
others it goes to infinity. If the multiplicity of the maximal coefficients is high then the leading term of the distribution
has a chi-square distribution with large degree of freedom what is approximately normal. Thus it is possible for the limit 
distributions for

-- Wasserstein distances for uniform distribution,

-- Wasserstein distances for standard normal distribution,

-- Ajtai distances for uniform distribution,

-- Ajtai distances for standard normal distribution,

\noindent
all are the normal one but the speed of the convergence is considerable slow. Using the marginal
quantile transformation one can ask what is the relation between the distances for standard normal
and uniform distributions. In the next table we give the basic statistics for Ajtai distances up to $4^8.$
Here $D=1$ stands for standard normal and $D=2$ for uniform distribution and the correlations come
from marginal quantile transformation. 

\bigskip
\begin{center}
{\bf {Table 3.}}
\end{center}
{\hspace{-1cm}{
\begin{tabular}{ccrrrrrrr}
  k & D & Sample Size&   Average  &   Stdev    &   Sequeness  & Kolmogorov &  Locus& Correlations\\
  1 & 1 &  42161 &   9.951413 &   5.606661 &   1.261530 &  16.717425 &  23626&  \\
  2 & 1 &  40184 &  19.597105 &   6.658678 &   0.877757 &  11.497639 &  22992&  \\
  3 & 1 &  40711 &  34.650952 &   7.678610 &   0.663583 &   8.510061 &  21130&  \\
  4 & 1 &  49367 &  58.719088 &   8.880549 &   0.507017 &   7.463537 &  26939&  \\
  5 & 1 &  20089 &  97.984671 &  10.266948 &   0.334140 &   3.379643 &   8789&  \\
  6 & 1 &  12988 & 163.848339 &  12.198350 &   0.277057 &   2.205296 &   4800&  \\
  7 & 1 &   5739 & 276.390900 &  14.609538 &   0.130521 &   0.986046 &   2560&  \\
  8 & 1 &   2333 & 471.869129 &  18.169085 &   0.130229 &   0.658314 &   1104&  \\
  1 & 2 &  42161 &   0.251985 &   0.154344 &   1.114916 &  14.115933 &  21684 &   0.891637  \\
  2 & 2 &  40184 &   0.655468 &   0.217523 &   0.539991 &   7.161248 &  21916 &   0.858883  \\
  3 & 2 &  40711 &   1.428708 &   0.294253 &   0.420951 &   5.765374 &  21836 &   0.872925  \\
  4 & 2 &  49367 &   2.519040 &   0.360076 &   0.433037 &   6.159368 &  29373 &   0.858270  \\
  5 & 2 &  20089 &   3.809825 &   0.400688 &   0.401140 &   3.978578 &  12257 &   0.807309  \\
  6 & 2 &  12988 &   5.210258 &   0.421534 &   0.390273 &   3.757436 &   7405 &   0.734815  \\
  7 & 2 &   5739 &   6.669811 &   0.431007 &   0.373776 &   2.231697 &   2844 &   0.645670  \\
  8 & 2 &   2333 &   8.138648 &   0.438881 &   0.363513 &   1.112873 &    898 &   0.558825  \\
\end{tabular}
}}

\bigskip\noindent
Because the number of points $4^8$ is rather large  in case of the Hungarian algorithm,
thus we reduced it to $4^5.$ In the next table we present the basic statistics for sample size $309.$

\bigskip

\begin{center}
{\bf {Table 4.}}
\end{center}

\begin{tabular}{crrrr}
Label & Sample type & Algorithm & Average & St.deviation \\ 
NH & normal sample &Hungarian algorithm & 39.184 & 5.453 \\
NA & normal sample& Ajtai algorithm & 97.497 & 10.485 \\
UA & uniform sample& Ajtai algorithm & 2.902 & 0.541 \\
UH & uniform sample &Hungarian algorithm & 1.756 & 0.451 \\
UB & uniform sample &adhoc improving & 2.224 & 0.500 
\end{tabular}

\bigskip\noindent
and the correlations are the followings:

\newpage

\bigskip
\begin{center}
{\bf {Table 5.}}
\end{center}
{\hspace{+3cm}{
\begin{tabular}{cccccc}
& NH & NA & UA & UH & UB \\
NH & 1.000 & 0.665 & 0.681 & 0.766 & 0.717 \\
NA & 0.665 & 1.000 & 0.569 & 0.469 & 0.530 \\
UA & 0.681 & 0.569 & 1.000 & 0.896 & 0.981 \\
UH & 0.766 & 0.469 & 0.896 & 1.000 & 0.942 \\
UB & 0.717 & 0.530 & 0.981 & 0.942 & 1.000 \\
\end{tabular}
}}
\bigskip

The correlation between Hungarian and Ajtai algorithms for uniform distribution is $0.896$
which is astonishingly high. It proves that the Ajtai algorithm is very efficient. Even it is
efficient concerning the average but surprisingly it is efficient in measuring the habits
of the sample for having good marriages. Now it is a standard observation for Wasserstein
couplings: the condition that for any finite set of marriages no reordering of the couples
could improve the sum of distances is sufficient. For large number of points the condition
is hard to check. We use the simplest case with two couples, it is labeled us ``adhoc improving''.
One can say that it is natural that its correlation with original Ajtai is as high as $0.981$
because our ad-hoc improving starts with the Ajtai coupling. But its correlation with
the optimum is also improved: it is now $0.94.$

\bigskip

\section {Discussion}

The original Ajtai algorithm uses numerical medians and linear transformations fitting the medians 
to the interval halving. First we apply the median of the whole set of first coordinates and imagine a
vertical line dividing the unit square accordingly. The number of points in the rectangles
are equal. Next we divide independently with appropriate horizontal lines the left hand side
and right hand side rectangles in such a way that the number of points in the four rectangles
should be equal. Roughly what happen in the four rectangles are independent from each other 
if it is accordingly scaled. This concept leads to a
dynamical equation similar to (\ref{eq:dynam}) but in this case the four terms are independent.
This type of recursion immediately  leads to a normal limit distribution. When we believed
that the limit distribution is {\it {not}} Gaussian we desperately struggled against such
kind of dynamics but now we are content that the limit distribution is Gaussian for both
cases of Ajtai and Wasserstein distance. Of course for Wasserstein
distance the situation is  more complicated. The Ajtai algorithm works for arbitrary
number of points, we used the power of $4$ only
 for didactical reason.

\includepdf{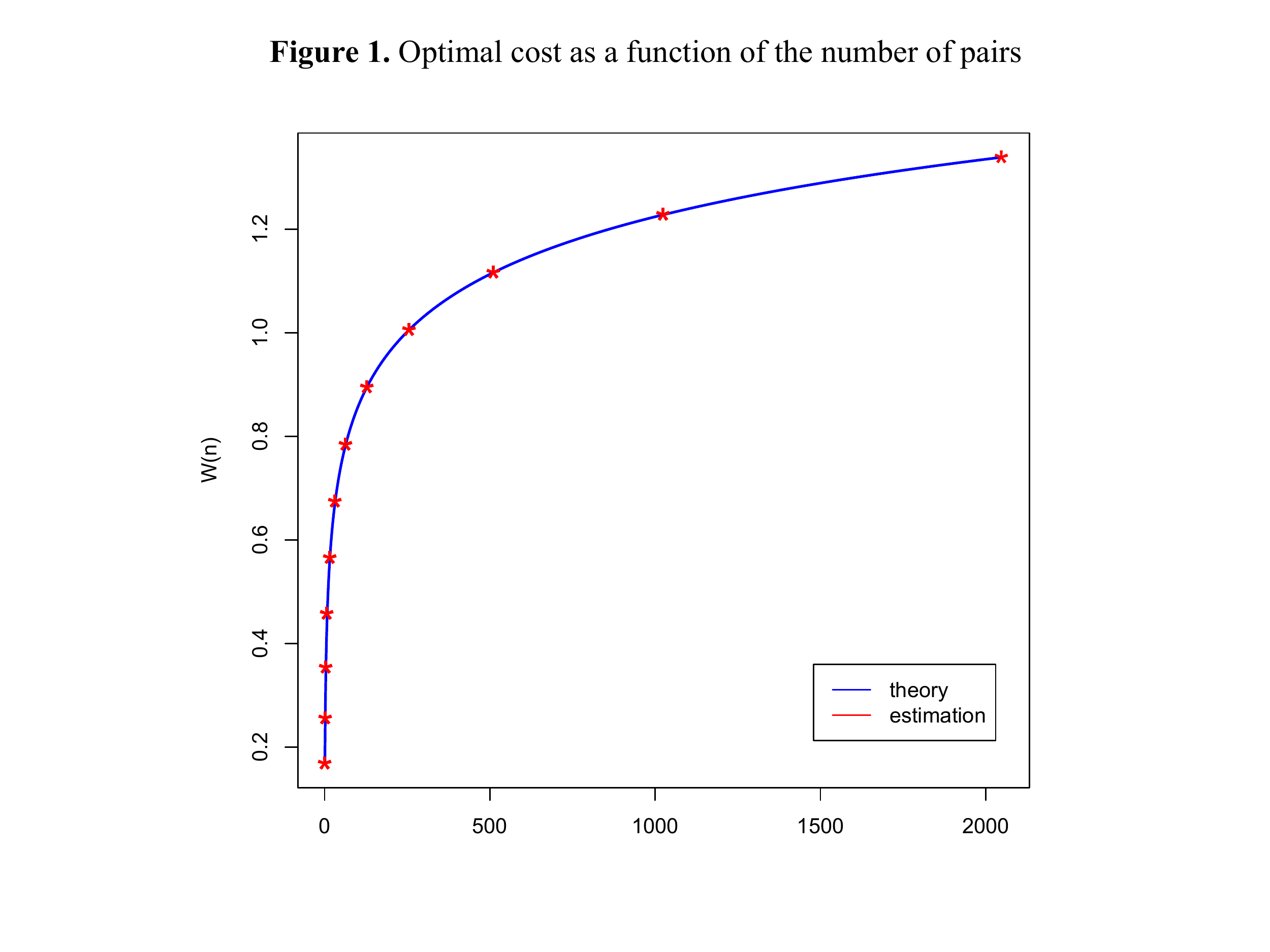}
\includepdf{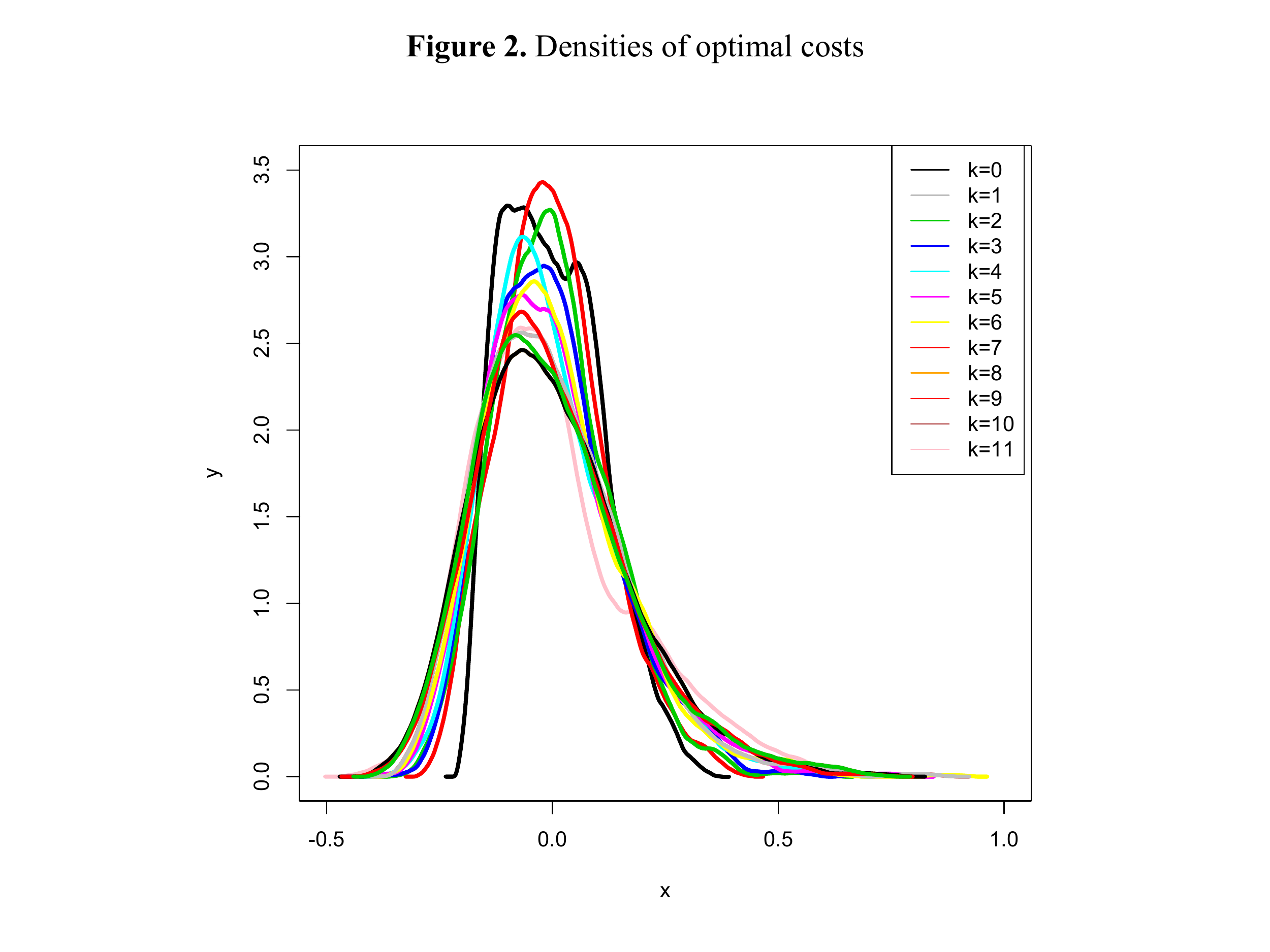}
\includepdf{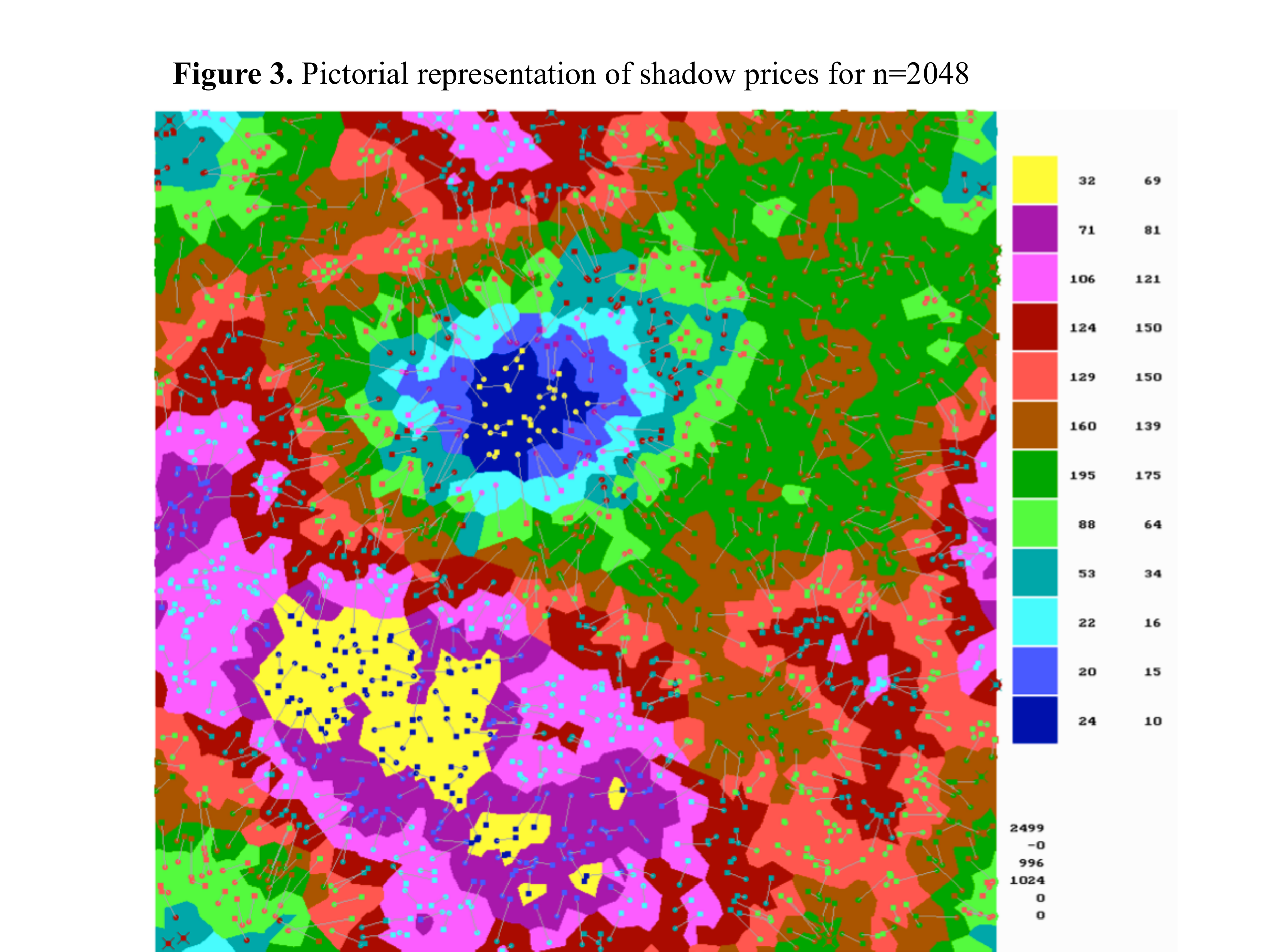}
\includepdf{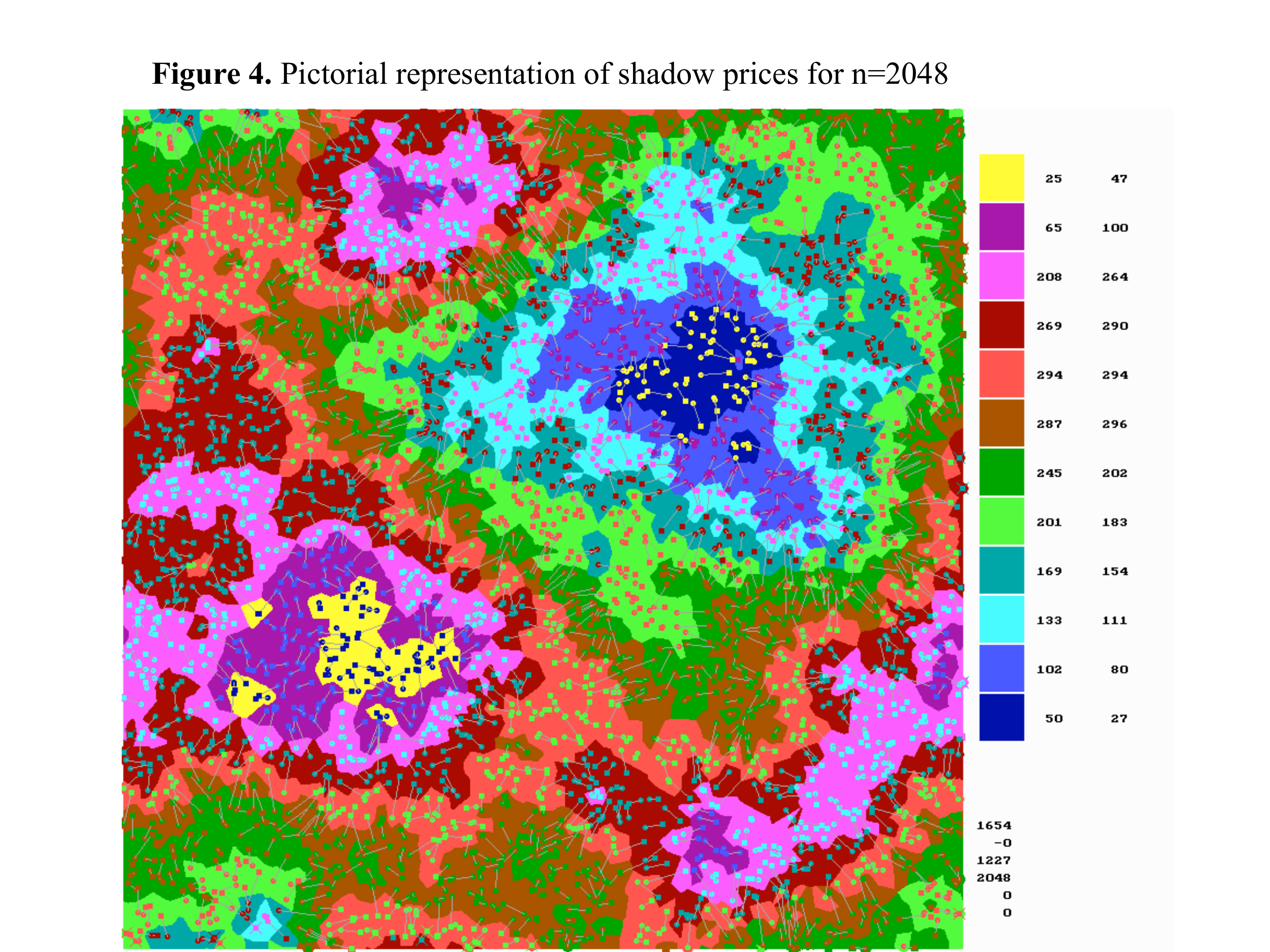}
\includepdf{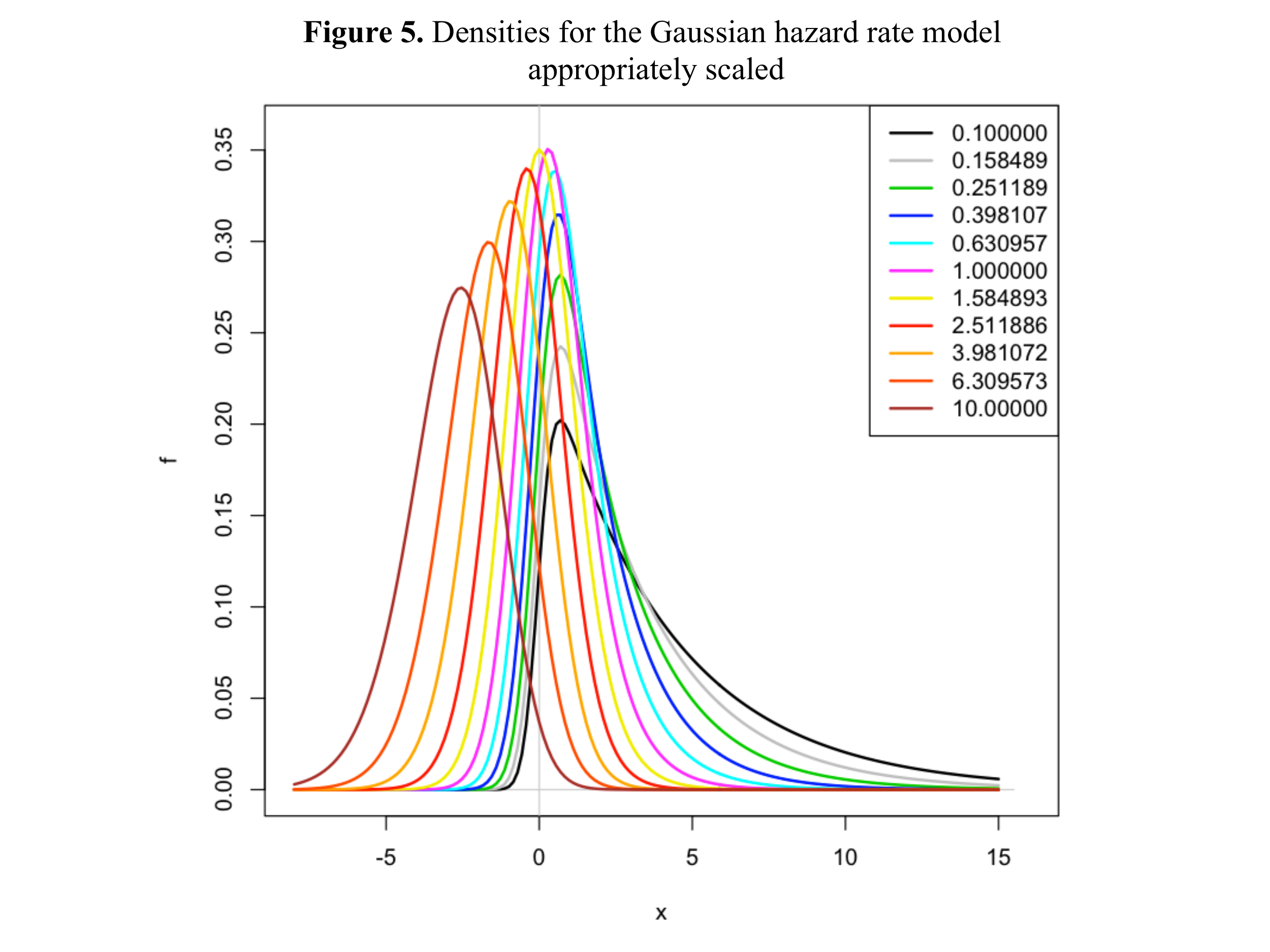}

\end{document}